# Distance-based subsidy rate design to incentivize ride-hail access to advanced air mobility hubs

Zhenglei Ji, Hai Yang, Joseph J. Y. Chow*

C2SMART Center, Department of Civil, Urban, & Environmental Engineering, New York University Tandon School of Engineering, 6 MetroTech Center, Brooklyn, NY 11201, USA

* Corresponding author: joseph.chow@nyu.edu

## Abstract

The success of advanced air mobility (AAM) operations is largely contingent on its effective integration with other ground transport modes. Under many use cases, AAM operators have to work with ride-hailing operators to create a seamless air taxi travel experience with adequate first and last-mile access. In investigating this multimodal coalition, this study proposes a distance-based subsidy rate design for AAM operators to incentivize ride-hail access to AAM hubs, incorporating air mobility operators' profitability considerations and travelers' route choices jointly. Using New York City (NYC) airport access as a case study, this study integrates high-volume for-hire vehicle (HVFHV) data from NYC taxi zones to consider real-world spatial demand distributions while considering passenger groups with different values of time (VOT) to derive insights on distinctive customer bases. Overall, the results show that AAM operators would need to subsidize the ride-hailing operators on vertiport access trips when air taxi operating costs exceed $12/mi. The analysis of ridership at AAM hubs indicates that ridership and profit contributions differ across different candidate vertiports in Manhattan, reflecting spatial demand heterogeneity. Additionally, having the airport access system in place, the taxi zones that generate the highest passenger demand to all three major NYC airports are identified under lower air taxi fare scenarios. These findings highlight how a distance-based subsidy rate design is beneficial in facilitating better access to vertiports and to foster high air taxi ridership with optimal AAM fare levels.

**Keywords:** advanced air mobility; urban air mobility; low-altitude economy; nelder-mead algorithm; bilevel programming

# 1. Introduction

The recent advancements in the advanced air mobility (AAM) sector are attributed to a combination of technological breakthroughs, capital investment, public policy support, and research activities (Cohen et al., 2021; Garrow et al., 2021; Goyal et al., 2018; Putnam & Littell, 2019; Riedel & Rozenkopf, 2022; Su et al., 2024). The use cases of AAM typically involve utilizing highly automated electric vertical takeoff and landing (eVTOL) aircraft in two domains: Urban Air Mobility (UAM) and Regional Air Mobility (RAM) (Johnson & Silva, 2022). In congested cities, UAM can offer mobility services such as airport access, emergency services, medical transport, and air tours (Mahmassani et al., 2024; Rath & Chow, 2022; Thipphavong et al., 2018). For RAM, eVTOL aircraft can be particularly useful for intercity travel, airport access in rural and remote communities, and reliable air connections to geographically isolated islands and mountainous regions (Antcliff et al., 2021; Cohen et al., 2024).

The rapid growth in the AAM sector is also evident in the progress made by various eVTOL manufacturers. Particularly in the United States, Joby, Archer Aviation, and Beta Technologies have made substantial progress in the aircraft type certification process with the Federal Aviation Administration (FAA) and have established manufacturing facilities across the country. Moreover, the U.S. Department of Transportation (USDOT, 2025) released a national strategy to institutionalize AAM as a national transportation priority, transforming it from a technology driven concept into a governed, coordinated, and long-term public policy initiative. Jonas et al. (2021) project the sector to reach a total addressable market (TAM) of $1 trillion globally by 2040 and $9 trillion by 2050. The sector's promising potential underscores the need for providing efficient mobility services in low-altitude airspace.

Although AAM services aim to offer an efficient, low-noise, and sustainable air transport option, eVTOL operations remain inherently reliant on designated takeoff and landing infrastructure (i.e., vertiports), whereas ride-hailing services typically involve longer travel times but provide direct door-to-door service. Fig. 1 depicts potential AAM use cases within New York City, where links labeled "AT" represent air transport legs and those labeled "GT" denote ground transport legs. Typical applications include trips by residents or visitors traveling from residential areas or hotels to destinations such as airports, hospitals, or stadiums, as well as intra-metropolitan travel within suburban areas, including trips between corporate campuses and manufacturing facilities. In most cases, these services would require initial ground access links to connect travelers to vertiports. This characteristic of AAM operations has motivated a growing body of research on its integration with other transportation modes (Yang et al., 2023; Zhang et al., 2025; Zhao & Feng, 2024; Zhao et al., 2025a; Zhao et al., 2025b), however, limited research has examined the network-level integration of AAM and ride-hailing services, particularly in the coordinated design of fares and subsidy mechanisms across multiple mobility operators to support the commercial viability of the emerging low-altitude economy.

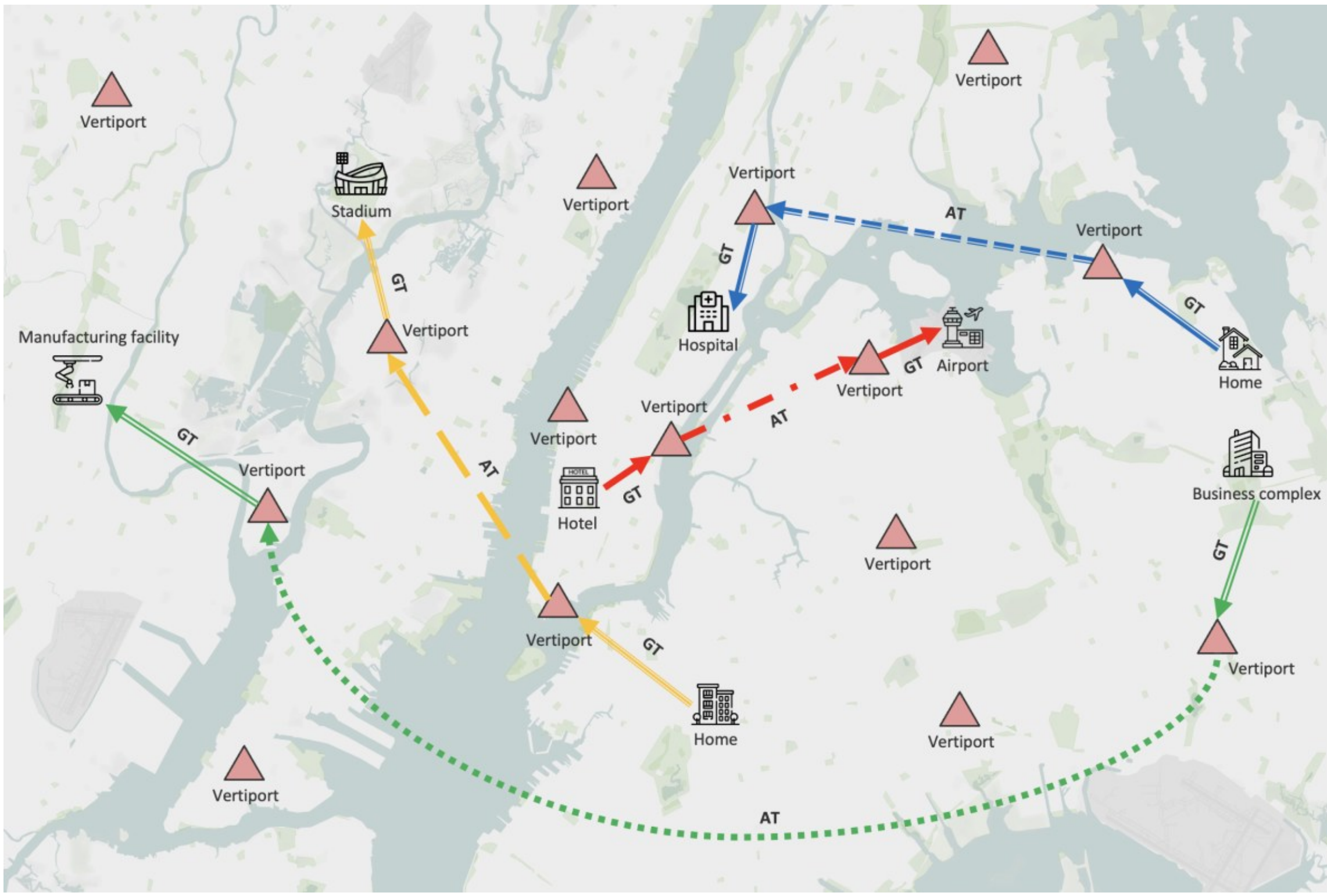

**Fig. 1.** Potential AAM use cases in NYC with hypothetical locations for origins, destinations, and vertiports.

One of the primary challenges facing AAM operations is uncertainty regarding its commercial viability, largely driven by the high operating costs associated with offering these services. Blade Air Mobility, a leading U.S. air mobility service provider, currently operates flights using conventional helicopters and has identified high operational costs as the main hurdle to profitability, a limitation that is expected to lessen with the adoption of eVTOL aircraft (Chokshi, 2025). The AAM operators have also undertaken efforts to integrate their services into the existing transportation network. For instance, Joby Aviation has partnered with Delta Air Lines to integrate eVTOL air taxi services into Delta's flight ticket booking systems for seamless city-to-airport trips, and more recently has advanced this commitment by bringing their services into Uber's widely used ride-hailing platform to create a more connected AAM travel experience (Joby Aviation, 2022, 2025). This shows that AAM's success depends not only on replacing helicopters for quiet and sustainable operations in low-altitude airspace, but also on being embedded into an integrated and coordinated mobility ecosystem.

This study proposes a modeling framework to determine optimal subsidy rates to travelers for ride-hail access that transforms vertiports into multimodal cyber-physical mobility hubs for AAM, where multiple modes operated by multiple operators make use of digital platforms to provide bundled services to travelers with transfers at physical hubs. Such systems require the use of game-theoretic models to design the incentives for cooperation. This study adopts an approach similar to that of Yang and Chow (2026), who developed a bilevel optimization model based on assignment games with transferable utilities to capture interactions between the mobility hub provider (leader) and the matched hub access providers and travelers (followers). The model incorporates stochastic path choices by mobility providers and travelers, based on the stochastic assignment game framework following Liu et al. (2025), but improves scalability by using a link-based representation of coalition choice based on the Perturbed Utility Route Choice (PURC) model (Fosgerau et al., 2022) for the lower-level assignment problem. However, the bilevel solution algorithm proposed by Yang and Chow (2026) is formulated for general mobility hub systems and subsidy design. In the case of AAM vertiport access, a simpler assignment game can be evaluated with the AAM operator serving as the platform, a single ride-hail company with a pre-existing fare rate as the access provider, and a single subsidy rate per unit distance as one of the decision variables. This system results in a model variant that is solved with a more efficient customized Nelder-Mead (NM) algorithm.

In the case study of AAM access to airports in New York City (NYC), public taxi data and demand models from Rath and Chow (2022) are leveraged to estimate parameters for the PURC-based coalition choice model in the lower level consistent with the earlier study. This enables a model that represents AAM airport access market for NYC consistent with the literature, in which insights are derived for optimal air taxi fare, subsidies rate and the value of different AAM hubs.

The remainder of the paper is structured as follows. Section 2 reviews the literature on multimodal integration of AAM operations, studies on assignment games formulated as bilevel optimization models, and identifies research gaps and the contributions of this study. Section 3 introduces the methodological framework, including the formulation of the proposed bilevel program, and the customized solution algorithm. Section 4 presents the NYC airport access case study, including the estimation of user and operator cost parameters, vertiport considerations, input assumptions, and model results. Section 5 concludes the paper's main findings and discusses directions for future research.

## 2. Literature review

### 2.1 Advanced air mobility multimodal integration

Mobility hubs are important in enabling the collaboration between different mobility services. Specifically, in the context of AAM operations, vertiports function as intermodal hubs that connect air taxi services with ground transportation modes by providing dedicated infrastructure for eVTOL takeoff, landing, and passenger transfer activities. The vertiport design, location, and capacity all influence AAM's accessibility, throughput, and public acceptance (EASA, 2021; FAA, 2021; Wu et al., 2025).

Accordingly, a number of studies have focused on how vertiports' placement and network design can influence multimodal connectivity. In a methodological approach similar to that utilized in this study, Zhang et al. (2025) used a bilevel program to model the land-air coordination for AAM, but they focused primarily on vertiport planning decisions aiming to minimize delay while considering congestion rather than the dynamics between multiple mobility service providers. Zhao and Feng (2024) focused on strategic vertiport planning for Beijing, considering adequate access to rapid transit and taxis, and found that a network of 19 optimally located vertiports could achieve an average reduction of about 80% in door-to-door multimodal travel time compared with ground only travel. Zhang and Wu (2021) developed a framework for designing an on-demand UAM operation network with optimal locations for vertiports and found that personal vehicles and ride-hailing services were the primary choices for access and egress at vertiports. Additionally, the study found that ground access and egress to vertiports account for roughly 70% of the total AAM trip service time, underscoring the importance of vertiport connectivity to AAM adoption.

Additionally, Tuchen (2020) focused on multimodal integration of AAM by developing a mode-agnostic data exchange architecture and illustrating how AAM can be dynamically coordinated with other modes to support seamless end-to-end trips, including ground transit, ride-hailing, rail, and commercial aviation. Although these studies provide insights into the collaboration between different operators from a vertiport planning perspective, they fail to capture the multi-operator aspect of AAM services.

Modeling multi-operator settings has been done in the context of Mobility-as-a-Service (MaaS) systems which integrate multiple transport modes into a unified two-sided market via a digital platform, capturing network effects between travelers and service providers while considering pricing, incentives, and service bundling decisions (Djavadian & Chow, 2017; Hensher & Hietanen, 2023; Xi et al., 2024). Modeling such integrated platforms requires models that capture the interactions of multiple operators in cooperation or coopetition (Bird, 1976; Derks & Tijs, 1985). Noncooperative game frameworks (Harker, 1988; Zhou et al., 2005) ignore the potential for multiple operators to share revenues. The broader MaaS ecosystem has been modeled using assignment games (Shapley & Shubik, 1971) as a cooperative game approach for matching travelers with mobility providers (Liu & Chow, 2024; Pantelidis et al., 2020; Rasulkhani & Chow, 2019; Yao & Zhang, 2024).

Several studies have explored how travelers might adopt air mobility within the multi-operator setting as part of a MaaS environment. Zhao et al. (2025a) introduced the concept of "Air Mobility as a Service (AMaaS)" examining commuters' willingness to incorporate air mobility legs into their daily travel under a MaaS paradigm. Using a stated-preference (SP) choice survey in Beijing, they found that there is a preference for subscription models that offer discounted bundles and travelers responded positively to the idea of multimodal UAM trips such as taking an air taxi to a transit hub, especially when incentives like government subsidy or lower fares were in place.

Another SP study by Zhao et al. (2025b) focused on vertiport choice for UAM and airline bundled trips. The results show that passengers prefer small, nearby vertiports integrated with public transport and offering adequate amenities. Vertiport access distance is found to be a significant influence on air mobility adoption, which reinforces the importance of first-mile/last-mile connectivity.

Berger (2023) investigated door-to-door AAM trips using an agent-based simulation framework that focused on evaluating the combined impacts of these multimodal trips on travel time, mode share, and energy consumption.

Although these studies have investigated travelers' preferences for mode choice, different multi-operator settings motivate a closer examination of strategic-level decisions among multiple operators, particularly with respect to pricing and subsidy decisions.

### 2.2 Assignment games as bilevel optimization models

Game theoretic approaches have been widely adopted to model both the deterministic and stochastic assignment games in a two-sided market between travelers and operators. This approach builds on the foundational work of Gale and Shapley (1962), who proposed a stable matching framework between buyers and sellers. Shapley and Shubik (1971) extended this framework by allowing utilities to be transferable between the two sides of the market, which allows for cooperative assignment games with stable matching where no individual agent or coalition of buyers and sellers can increase their utility by deviating from the assigned matching.

In particular, the deterministic assignment game models for MaaS markets assume homogeneous travelers within each population segment with identical utilities and behavioral responses to costs and prices (Rasulkhani & Chow, 2019; Pantelidis et al., 2020; Liu & Chow, 2024). However, these deterministic assignment game models often rely on perfect information, binary matching variables, and homogeneous traveler groups. As a result, the algorithms for such models are difficult to scale because of the challenge of solving mixed-integer nonlinear programming problems. Liu et al. (2025) sought to overcome this challenge with a stochastic assignment game framework where travelers and mobility providers collectively choose paths as coalitions to form. The framework becomes a bilevel problem with a lower-level stochastic assignment (Fisk, 1980) and an upper-level platform pricing decision.

Whereas the deterministic assignment game can have multiple cost allocations corresponding to an assignment or have no feasible allocations requiring additional interventions, the stochastic assignment game has a one-to-one relationship between the flow assignment and the cost allocations and is guaranteed to be non-empty. These properties make the model more desirable. The path-based choice set generation remains a scalability challenge in that study.

Yang and Chow (2026) overcome that challenge by adopting a link-based equivalent formulation that reflects the assignment outcome of coalition choice based on the PURC model from Fosgerau et al. (2022) and further results in a convex lower-level model. They apply the framework to the design of mobility hubs to connect rail hubs with on-demand mobility providers for first/last mile access, re-interpreting the pricing decisions as subsidies to last mile mobility operators while charging access fees to travelers. A gap function-based exact algorithm is developed to solve the integrated single level formulation with Karush-Kuhn-Tucker (KKT) conditions for the lower-level model. However, the convergence speed can depend on the variations of the problem setting being considered. No variation of the model has been customized for AAM hubs.

### 2.3 Research gap and contributions

An effective incorporation of AAM into urban transportation systems requires achieving coordination with ground transport operators to facilitate an integrated travel experience. Although there have been many studies on vertiports' role in AAM multimodal integration, they tend to focus on its optimal location, flight capacity, and service coverage. While these works demonstrate the benefits of optimal AAM hub locations, they do not adequately capture the multimodal collaboration between the different operators. Furthermore, while some studies highlight the benefits of providing AAM service through bundled trips, subscription models, and multimodal vertiport access trips, there have yet to be studies specifically on the collaboration between AAM and ride-hailing operators on subsidizing vertiport access trips. Additionally, from a methodological perspective, cooperative and noncooperative game frameworks have been applied in MaaS contexts (Bird, 1976; Liu & Chow, 2024; Rasulkhani & Chow, 2019; Shapley & Shubik, 1971; Yang & Chow, 2026) but the application to AAM operations has been limited.

In summary, a research gap exists in the design of effective incentive mechanisms to facilitate seamless integration between AAM and ride-hailing operators. Furthermore, there is a lack of studies focused on identifying optimal fare and subsidy structures that encourage ride-hailing operators to provide first and last-mile connectivity to AAM hubs.

To address these gaps, our work makes the following contributions:

- The study adapts a game-theoretic framework to AAM-specific use cases for scalable networks and solves it using an efficient NM algorithm. The model captures the interactions between the operators and travelers to provide a decision-support tool for operating air mobility services.
- The study proposes a framework to determine optimal distance-based subsidy rates for ride-hail access to vertiports, enabling them to become multimodal cyber-physical AAM hubs where multi-operator trips can be coordinated through digital platforms operated by the AAM provider subsidizing local ride-hail providers.
- The study implements an NYC airport access case study, leveraging actual demand and geospatial data to generate realistic scenarios to evaluate the dynamics between the AAM provider, the taxi operators, and travelers.

## 3. Methodology

The problem is formulated as a bilevel program between operators (leaders) and travelers (followers). This section first includes the upper and lower-level formulations, then it introduces a customized NM solution algorithm, and presents a performance comparison to demonstrate its efficiency. The notations used in Section 3 and Section 4 are included in Table 1.

**Table 1**

Notations

| Sets | |
|---|---|
| $L_v$ | Set of vertiport access links |
| $L_t$ | Set of vertiport transfer links |
| $L_g$ | Set of direct ground links to the airports |
| $L_c$ | Set of links from centroid to MOD zones |
| $L_a$ | Set of air taxi links |
| $S$ | Set of origin destination (OD) pairs and segments |
| **Parameters** | |
| $f_g$ | Ground taxi fare rate (USD per mile) |
| $c_a$ | Air taxi operating cost rate (USD per mile) |
| $c_g$ | Ground taxi operating cost rate (USD per mile) |
| $c_v$ | Vertiport access operating cost rate (USD per passenger) |
| $d_l$ | Length of link $l$ (miles) |
| $t_l$ | Travel time on link $l$ (minutes) |
| $\beta_s$ | Value-of-time coefficient for segment $s$ (USD per minute) |
| $f_a^{max}$ | Maximum allowable air taxi fare (USD per mile) |
| $\delta_{i,l}$ | Incidence coefficient for node $i$ and link $l$ |
| $b_{i,s}$ | Net flow requirement at node $i$ for OD pair and segment $s$, set to -1 for a source node, +1 for a sink node, and 0 else |
| $Q_s$ | Number of passengers for every OD pair and segment $s$ |
| $\kappa$ | Operating cost per passenger (USD) |
| $Z_l$ | Max capacity for vertiport dummy links (number of passengers) |
| $\alpha_1$ | Weighting factor for user costs |
| $\alpha_2$ | Weighting factor for operator costs |
| Decision Variables | |
| $f_a$ | Air taxi fare rate (USD per mile) |
| $r$ | Subsidy rate to travelers on all vertiport access links (USD per mile) |
| $x_{l,s}$ | Passenger flow on link $l$ for OD pair and segment $s \in S$ |
| $v_l$ | Passengers for vertiport dummy link $l$ (number of passengers) |

### 3.1. Model specification

Bilevel programming is a framework designed to capture hierarchical decision-making processes between two parties. In this structure, an upper-level (leader) problem and a lower-level (follower) problem are jointly formulated, where the leader's decisions both influence and are constrained by the follower's responses. This design forms a Stackelberg game, in which the leader acts first and the follower subsequently responds by optimizing its objective given the leader's decision. The method has been widely applied in transportation research to model interactions between stakeholders like operators, regulators, and travelers, allowing for the analysis of competition, pricing decisions, and network design.

We adopt a framework based on Yang and Chow (2026) and apply it in the air mobility case in NYC, where a link-based approach is considered to eliminate the need for path enumeration and increase the scalability of the method. Furthermore, this approach also allows one to decompose the overall journey into smaller segments that have different operators to investigate the partnership between ride-hailing operators and AAM operators. The following modifications to the formulation from Yang and Chow (2026) are made, divided to upper- and lower-level portions.

The network structure makes use of dummy links for the vertiports connecting the ground and air subnetworks as separate layers. Travelers are divided into OD pairs and segments, defined as the set $S$, with demand level $Q_s$. Links belong to either the air routes $L_a$, ground access routes $L_v$, vertiport dummy links $L_t$, direct ground links between origin and destination $L_g$ for travelers choosing not to use the AAM service, and centroid connector links $L_c$. In both Fig. 2(a) and Fig. 2(b), $l_1, l_2 \in L_a$ represent air taxi links (blue dashed arrows), while $l_3, l_4 \in L_g$ denote the vertiport access links (black solid arrows) operated by ride hailing providers. The vertiport dummy links are achieved by splitting each vertiport node $i \in N_v$ into separate dummy nodes. Specifically, a single vertiport node $i$, as seen in Fig. 2(a), is converted to two dummy nodes, $i^a$ (air layer) and $i^g$ (ground layer), connected by dummy links $l_5, l_6 \in L_t$ in each direction, as depicted in Fig. 2(b).

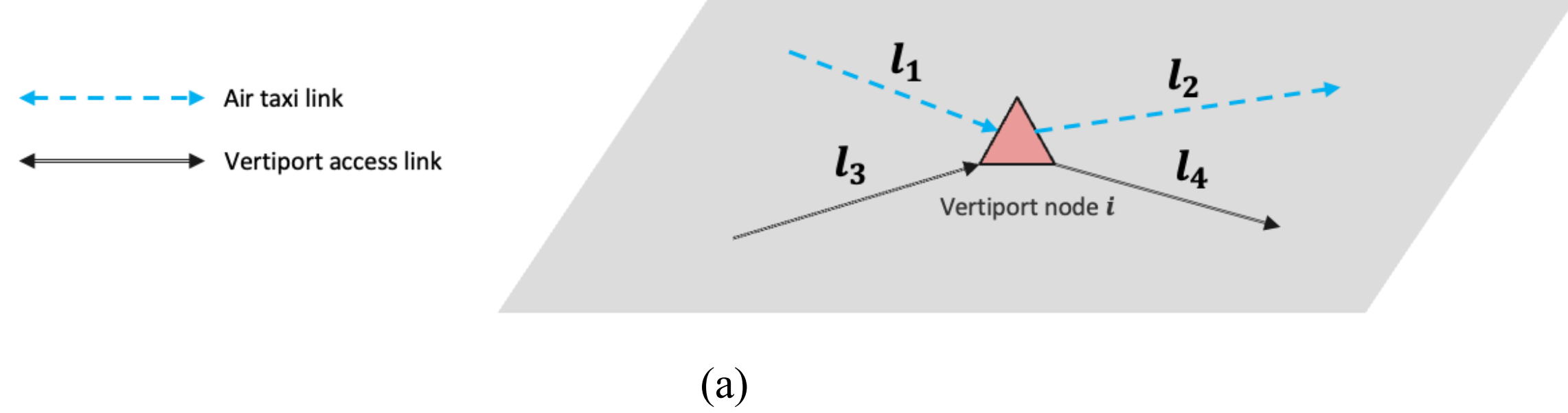

(a)

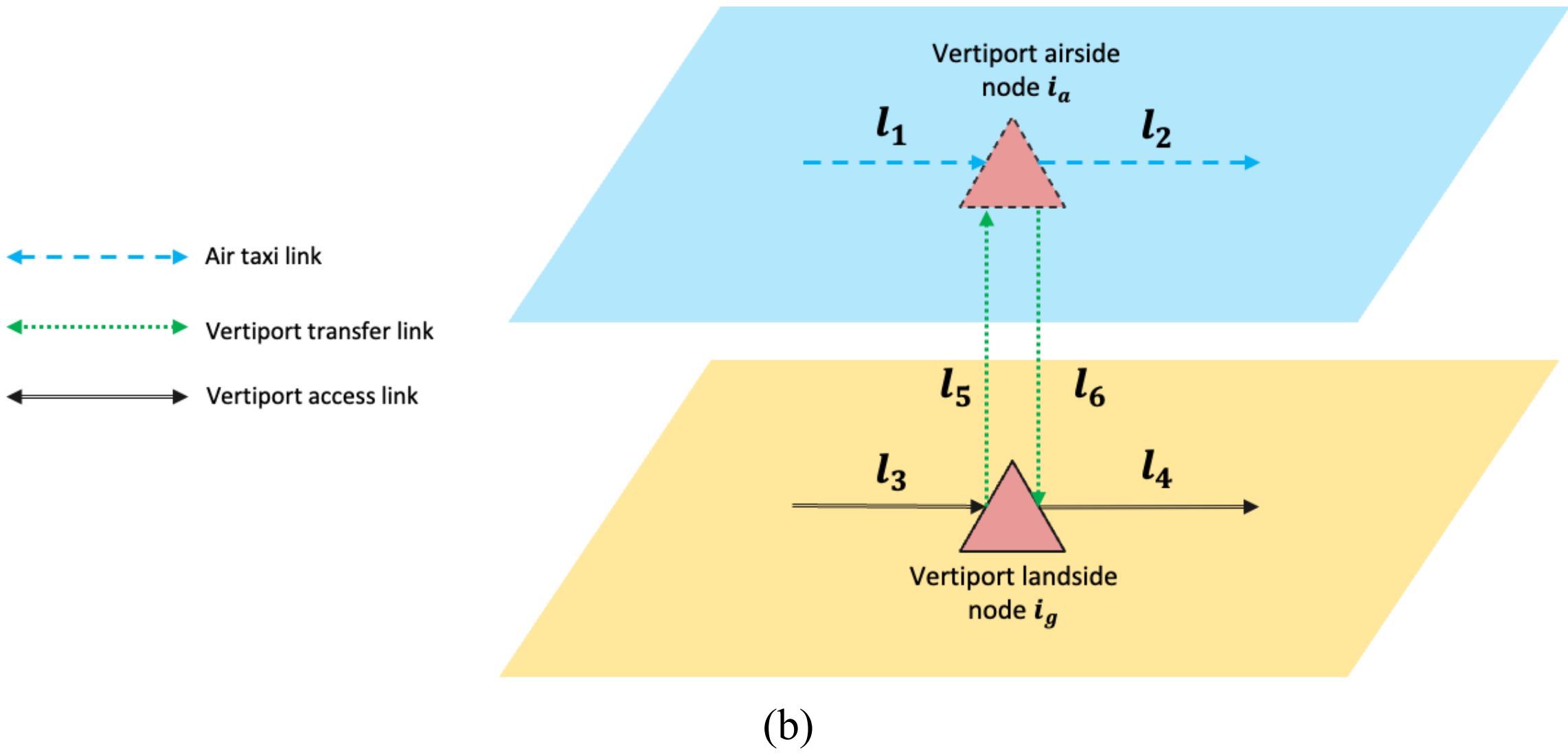

(b)

**Fig. 2.** Illustration of converting of (a) a single vertiport node $i$ with air taxi links $l_1, l_2 \in L_a$ and vertiport access links $l_3, l_4 \in L_v$ into (b) separate airside and landside dummy nodes $i^a$and $i^g$ with air taxi links $l_1, l_2 \in L_a$, vertiport access links $l_3, l_4 \in L_v$, and transfer link $l_5, l_6 \in L_t$ connecting $i^a$and $i^g$.

### *3.1.1 Lower-level formulation*

The lower level captures the joint behaviors of the travelers and the operators (both AAM and access ride-hail providers), determining both the traveler flows and the capacities assigned by the operators to the links. The objective Eq. (1) represents the perceived total system payoff $\Phi_1$ with dispersion effect under a stochastic assignment (see Yang and Chow, 2026). In a stochastic assignment, the heterogeneity in perceived cost parameters can lead to a dispersed set of probabilistic flows, i.e., stochastic assignments, instead of a single deterministic optimum. Eq. (1) is divided into three portions: the first two portions capture the traveler and the ride-hail operator perceived costs, respectively; the third portion captures the degree of dispersion under the PURC model framework.

The first term is scaled by $\alpha_1$, a perception weight for traveler cost that should be estimated from flow data. The term $\beta_s t_l$ represents travel time disutility for passenger segment $s \in S$ with value of time $\beta_s$. For vertiport access links $L_v$, the per-mile cost is $(f_g - r)$, where $f_g$ is an exogenous ride-hailing fare and $r$ is the optimized subsidy rate, multiplied by link distance $d_l$. Direct ground $L_g$ and centroid connector links $L_c$ are subject to a baseline ground fare per-mile $f_g$, while air-taxi links $L_a$ apply the air taxi per-mile fare $f_a$. When multiplied by the link percent flows $x_{l,s}$ and weighted by $\alpha_1$, the entire expression captures the total perceived generalized traveler cost. The values of time $\beta_s$ are either estimated from data or assumed.

$$\min_{x_{l,s}\, v_l} \Phi_1 = \sum_{s\in S} Q_s \Bigg( \alpha_1 \Bigg( \sum_{l\in L_v} [\beta_s t_l + (f_g - r) d_l] x_{l,s} + \sum_{l\in L_g \cup L_c} [\beta_s t_l + f_g d_l] x_{l,s} + \sum_{l\in L_a} [\beta_s t_l + f_a d_l] x_{l,s} \Bigg) + \alpha_2 \Bigg( \sum_{l\in L_a} c_a d_l x_{l,s} + \sum_{l\in L_c \cup L_g} (c_g - f_g) d_l x_{l,s} + \sum_{l\in L_t} \kappa v_l + \sum_{l\in L_v} c_v v_l \Bigg) + \sum_{l\in L} d_l \big(x_{l,s}\big)^2 \Bigg) \tag{1}$$

$$s.t.$$

$$\sum_{l\in L} \delta_{i,l} x_{l,s} = b_{i,s}, \quad \forall i \in N_v, \forall s \in S \tag{2}$$

$$\sum_{s\in S} Q_s x_{l,s} \le Z_l v_l, \quad \forall l \in L_t \cup L_v \tag{3}$$

$$\sum_{s\in S} Q_s x_{l,s} \le Z_l, \quad \forall l \in L_c \tag{4}$$

$$v_l \in [0,1], \quad \forall l \in L_t \tag{5}$$

$$x_{l,s} \in [0,1], \quad \forall l \in L, \forall s \in S \tag{6}$$

The second part of Eq. (1) is scaled by $\alpha_2$, the weight for operator costs. These include per distance operating cost rates via air $c_a$ or ground $c_g$, the cost $\kappa$ of operating each vertiport at capacity $v_l$, $l \in L_t$, and cost $c_v$ of serving an access route at capacity $v_l, l \in L_v$.

The final component of the objective function is the link-based perturbed utility term $d_l\big(x_{l,s}\big)^2$ adopted from Fosgerau et al. (2022) to capture dispersion effect. The greater this value relative to $\alpha_1$ and $\alpha_2$, the more indifferent the coalitions are to the expected costs. Different functions may be used as long as they feature strict convexity and the first and second derivatives are zero at zero. Here, $\big(x_{l,s}\big)^2$ is chosen as one of two recommended functions by Fosgerau et al. (2022) and shown to be effective for stochastic hub assignment games by Yang and Chow (2026).

Eqs. (2) are flow conservation constraints, Eqs. (3) – (4) are capacity constraints, and Eqs. (5) – (6) are boundary constraints.

The lower-level problem is a quadratic program (QP) with only a single quadratic term ($\sum_{l\in L} d_l\big(x_{l,s}\big)^2$) and well-defined solution space bounded between 0 and 1 for all the decision variables. This makes it fairly easy to solve with commercial QP solvers.

### *3.1.2 Upper-level formulation*

In the upper-level problem, the AAM provider maximizes overall profit $\Omega_0$ by designing the air fare and ride-hail subsidy for the travelers in objective function Eq. (7), where the AAM provider is also the platform operator. The overall objective function value equals the profit collected by the AAM provider across the network minus the total subsidy given out to ride-hail access travelers. To focus on the subsidy design, the AAM fare rate $f_a$ and subsidy rate $r$ are both treated as scalar network-wide values. This difference in the model formulation from Yang and Chow (2026) is deliberate, as the smaller dimensionality allows for even faster global solution algorithms that still reflect realistic design variables for more insightful analysis of subsidies.

$$\max_{f_a, r} \Omega_0 = \sum_{l \in L_a} \sum_{s \in S} (f_a - c_a) d_l Q_s x_{l,s} - \sum_{l \in L_v} \sum_{s \in S} r d_l Q_s x_{l,s} \tag{7}$$

$$s.t.$$

$$\sum_{l \in L_a} \sum_{s \in S} f_a d_l Q_s x_{l,s} \geq \sum_{l \in L_a} \sum_{s \in S} c_a d_l Q_s x_{l,s} + \sum_{l \in L_v} \sum_{s \in S} r d_l Q_s x_{l,s}, \quad \forall l \in L_a, \forall l \in L_v, \forall s \in S \tag{8}$$

$$0 \leq f_a \leq f_a^{max} \tag{9}$$

$$(x, v) = \arg\min \Phi_1(x, v; f, r) \tag{10}$$

The objective Eq. (7) maximizes revenue minus the subsidy costs and vertiport operating costs. Eq. (8) ensures that the AAM operator is profitable. Eq. (9) bounds the air fare. Eq. (10) is the lower-level problem.

### 3.2 Solution algorithm

By structuring the upper-level pricing problem into a two-dimensional distance-based subsidy and air fare rates, we can take advantage of much faster algorithms to solve the problem than the more generic method proposed in Yang and Chow (2026). We propose solving the bilevel problem using a variant of the NM algorithm, a derivative-free direct search method originally introduced by Nelder and Mead (1965). Its primary benefit is that it only requires function evaluations, making it effective for low-dimensional problems where gradients are computationally expensive to obtain. Such is the case where the upper-level problem is low-dimensional while the lower-level problem can be very high-dimensional. The method uses a simplex (polytope) with $n + 1$ vertices in $\mathbb{R}^n$ space and iteratively updates it through operations like reflection, expansion, contraction, and shrinkage operations to converge toward a local minimum.

However, NM is not inherently designed for constrained optimization, as its updates rely only on function value comparisons and do not enforce feasibility constraints. Without modifications, it may generate infeasible solutions, requiring penalty or barrier methods to handle constraints. To address this limitation, Price (2021) introduced the Pseudo-Expand Nelder-Mead for Constrained Optimization (PENMECO) method, which incorporates an interior-point barrier to enforce feasibility and ensures convergence to KKT consistent solutions without requiring gradient information.

**Table 2**
Hyperparameters used in Section 3.2.

| | |
|---|---|
| $W$ | Barrier update scheduling constant |
| $w_0$ | Initial barrier weight |
| $h_0$ | Initial polytope step size |
| $p$ | Barrier parameter |
| $\tau$ | Shrink factor for step size for barrier weight |
| $\rho$ | NM reflection coefficient |
| $\nu$ | NM contraction coefficient |
| $\eta$ | NM expansion coefficient |
| $h_{min}$ | Minimum allowable step length |
| $N_{max}$ | Maximum number of barrier objective function evaluations |
| $z$ | Decision vector for NM search |

Consider an objective function $f(x)$, where $c_i(x)$ denotes the $i$-th inequality constraint and $k$ is the total number of constraints in Eq. (11).

$$\min_{x \in R^n} f(x) \quad \text{s.t.} \quad c_i(x) \leq 0,\ i = 1, \ldots k \tag{11}$$

The Barrier function is described as Eq. (12), where $w > 0$ is the barrier weight and $\epsilon > 0$ is a small quadratic tie-breaking parameter. This allows the model to enforce the constraints to derive only feasible solutions.

$$B(x, w) = f(x) + w \sum_{i=1}^{q} P\big(c_i(x)\big) + \epsilon \|x\|^2 \tag{12}$$

The penalty component $P(c_i)$ in $B(\cdot)$ adopts the logarithmic barrier proposed by Price (2021), as defined in Eq. (13). This formulation ensures differentiability and strict convexity over the feasible region, thereby enforcing interior feasibility. These properties guarantee that the barrier function never underestimates the objective and that feasibility is maintained throughout the optimization process.

$$P(c_i) = -\ln\left(\frac{-c}{p - c}\right), \qquad p > 0 \tag{13}$$

The overall implementation of the proposed solution procedure is summarized in Algorithm 1. A set of candidate solutions $(f_a, r)$ are generated for the upper-level problem $\Omega_0$, where $f_a$ represents the air-taxi fare (\$/mi) and $r$ denotes the subsidy rate (\$/mi). The candidate with the smallest constraint violation is selected as the initial point. A local coordinate search with step-size reduction is then applied to refine this point and identify a strictly feasible starting vector for the subsequent direct search process. At each iteration, the simplex is perturbed through reflection, expansion, contraction, or shrinkage operations to generate a new decision vector $z = (f_a, r)$. This updated vector is then used to solve the lower-level problem, where the traveler route choice response is re-evaluated. The lower-level problem is solved using Gurobi to global optimality. The resulting objective and constraint values are used to compute the barrier-adjusted upper-level objective function.

**Algorithm 1: Nelder-Mead barrier solution method for the constrained subsidy-enabled air mobility network problem**

1. **Input:** Link flows $x_{l,s}$ obtained from the lower level model and the hyperparameters.
2. **Initialization:** Generate a grid of candidate $(f_a, r)$ values and select feasible $z_0 \leftarrow (f_{a,0}, r_0)$
3. Set initial barrier weight $w \leftarrow w_0$ and initial step size $h \leftarrow h_0$
4. Construct initial polytope centered at $z_0$ and evaluate barrier objective at polytope vertices with the barrier function defined below
$$B(f_a, r, w) = -\Omega(f_a, r) + w \sum_i P\big(c_i(f_a, r)\big) + \epsilon \|f_a, r\|^2, \quad where\ P(c) = -\ln\left(\frac{-c}{p-c}\right)$$
5. Solve the lower-level model for each OD pair $s$ to obtain the flows $L_a$, $L_v$, $L_d$ and $L_t$
6. Construct initial simplex around $z_0$, and evaluate $B(\cdot)$ at vertices
7. **repeat**
8. Sorting: Sort the vertices by their barrier objective values $B(z_i)$ compute the centroid of the two best vertices $\tilde{z} \leftarrow \frac{1}{2}(z_1 + z_2)$
9. Reflection: $z_r \leftarrow \tilde{z} + \rho(z_{best} - z_{worst})$, evaluate $B(z_r, w)$
10. **if** $B(z_r) < B(z_{best})$ **then**
11. **accept** $z_r$
12. **else if** expansion
13. $z_e \leftarrow \tilde{z} + \eta(z_r - \tilde{z})$
14. **if** $B(z_e) < B(z_{best})$ **then**
15. **accept** $z_e$
16. **else if** contraction
17. $z_c \leftarrow \tilde{z} - v(z_{worst} - \tilde{z})$
18. **if** $B(z_c) < B(z_{best})$ **then**
19. **accept** $z_c$
20. **else**
21. **keep** $z_{best}$
22. **end if**
23. **end if**
24. Reshape polytope if stagnation, pseudo expand:
$$y_{mid} = \frac{1}{2}(z_{second-best} + z_{worst}), \qquad z_{pe} = y_{mid} + \frac{\eta}{\rho}(z_{best} - y_{mid})$$
25. **if** $B(z_{pe}) < B(z_{worst})$ **then**
26. **replace** $z_{worst}$ with $z_{pe}$
27. **else**
28. **return** quasi-minimal frame
29. **end if**
30. Update barrier weight:
31. **if** $w \sum_i \left(\frac{\Delta c_i}{max\,(|c_i(z_{best})|, \epsilon_c)}\right)^2 \leq W h^{3/2}$ then
32. $w \leftarrow \tau w$
33. **end if**
34. Terminate if $h < h_{min}$ or evaluation budget exceeded $N_{max}$
35. **Output:** Return best feasible $(f_a, r)$.

The barrier weight $w$ is reduced adaptively under the condition $w \sum_i \left(\frac{\Delta c_i}{max\,(|c_i(z_{best})|, \epsilon_c)}\right)^2 \leq W h^{3/2}$, where $\Delta c_i$ denotes constraint variation and $h$ represents the simplex size. Here, $\Delta c_i$ is computed across the current simplex as: $\Delta c_i = \max_{j \in V_m} \big| c_i(z_j) -$

$c_i(z_{best})\big|$, where $V_m$ is the set of simplex at iteration $m$ and $z_{best}$ is the vertex with the smallest barrier objective value. This ensures that $w$ decreases only when the simplex has contracted sufficiently, allowing the barrier term to tighten gradually while maintaining near constraint boundaries.

The barrier parameter $p$ regulates the curvature of the logarithmic penalty function, whereas $W$ controls the update rate of the barrier weight $w$ to maintain stability during contraction. A per-constraint barrier term is applied to penalize infeasibility as the solution approaches constraint boundaries. The newly assigned link flows are then computed again in the upper level solved using NM search. The polytope at every iteration is assessed using the barrier objective $B(f_a, r, w)$ which acts as a buffer, ensuring the final solution remains strictly within the feasible region. The algorithm terminates when the simplex size $h$ falls below $h_{min}$ or the evaluation limit $N_{max}$ is reached.

### *3.2.1. Algorithm runtime comparison*

To assess the computational performance of the proposed Algorithm 1, we compare it against the gap function approach used by Yang and Chow (2026) across three network sizes. The gap function method provides a more generalized way to solve the PURC-based stochastic assignment game by transforming the bilevel problem into a single-level problem. The comparison between the two algorithms is done on grid networks featuring randomly generated nodes and OD pairs as illustrated in Fig. 3.

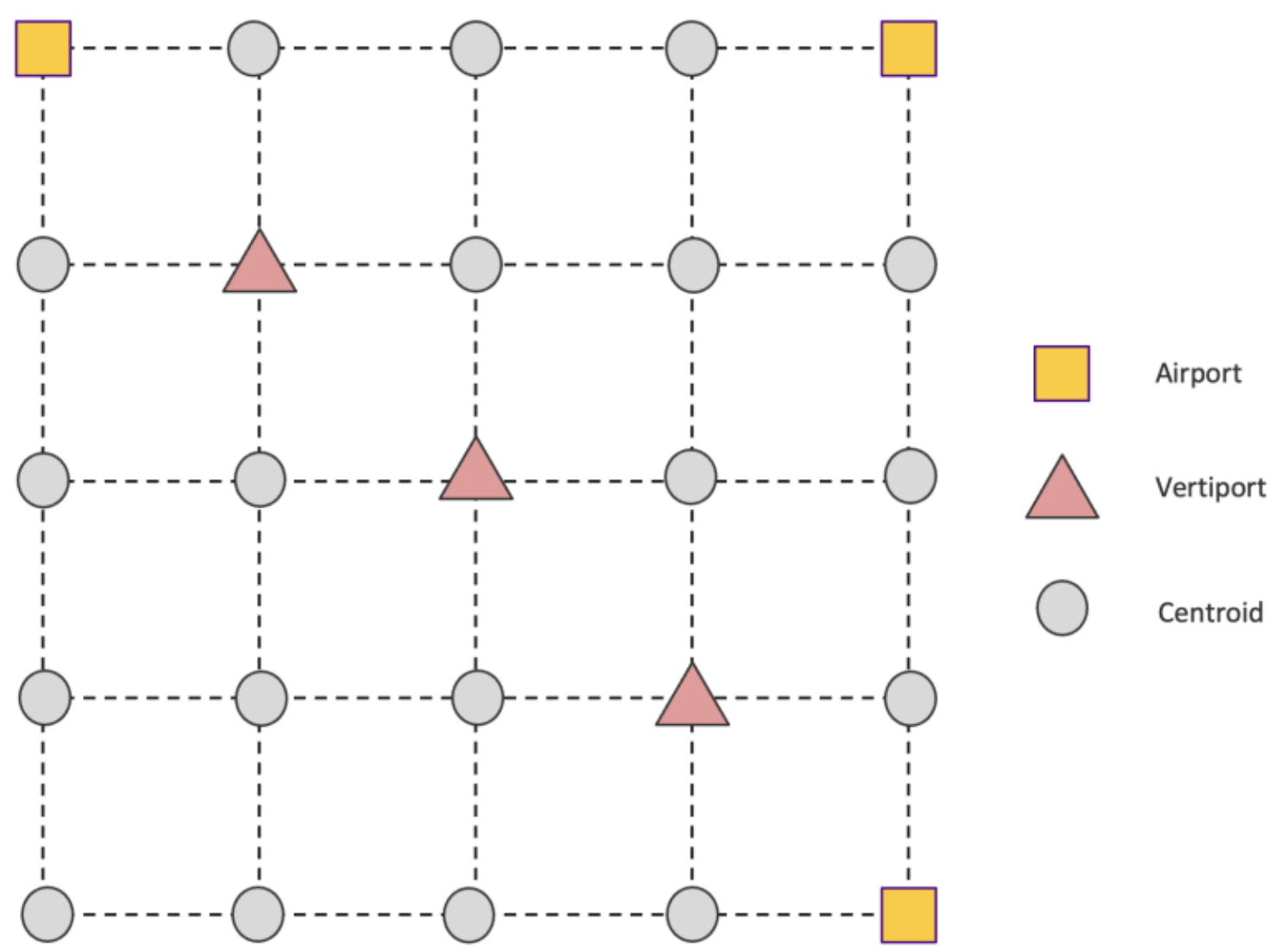

**Fig. 3.** Toy network tested for algorithm performance comparison (25 nodes, 4 OD pairs).

The tested networks include grid networks with 25, 100, 400, and 900 nodes, with the 25-node network example shown in Fig. 3. In each grid network, there are three airports that are generated at three of the corners, while vertiports and population zones are randomly placed. For each OD pair, ground and air travel times were computed as the sum of link-level times along the lattice (for ground) and Euclidean distances between vertiport-airport pairs (for air). Travel demand for each OD pair was drawn uniformly between 80 and 600 passengers per day to emulate variation in airport access demand. The randomly generated dataset is made available on Zenodo (Ji, 2025).

**Table 3**
Runtime performance of algorithms in reaching optimality across different network sizes.

| Nodes | OD pairs | Method | Runtime (s) *(%)* |
|---|---|---|---|
| 25 | 4 | Gap function | 0.13 |
| | | Algorithm 1 | 0.33 *(254%)* |
| 100 | 16 | Gap function | 85.66 |
| | | Algorithm 1 | 5.48 *(6%)* |
| 400 | 64 | Gap function | 751.47 |
| | | Algorithm 1 | 25.06 *(3%)* |
| 900 | 256 | Gap function | 4531.11 |
| | | Algorithm 1 | 136.98 *(3%)* |

All four grid networks were successfully solved to optimality using both the gap function method and Algorithm 1, as summarized in Table 3. The experiments were conducted under the model specification outlined in Section 3.1, with parameter settings outlined in Section 4.1.3 to ensure comparability across network sizes.

Both algorithms were implemented under identical model conditions. For the smallest 5×5 network (25 nodes, 4 OD pairs), both methods achieved convergence efficiently, with the gap function method slightly faster. However, as the network size and number of OD pairs increased, the runtime of the gap function method grew rapidly, exceeding 4500 seconds for the 30×30 network (900 nodes, 256 OD pairs). By contrast, Algorithm 1 scaled much more efficiently, solving the same instance in under 140 seconds (~ 3% of computation time of the gap function). These findings highlight that while the gap function algorithm is effective in solving instances with large decision space, its computational cost grows significantly with network size, whereas Algorithm 1 demonstrates superior scalability and robustness in handling much larger networks when restricting the upper level decision variables to two dimensions.

# 4. Case study: NYC airport access

## 4.1 Case study setup

The study considers airport access in NYC for our case study since it is a highly congested city where using low-altitude airspace for transportation is ideal. In addition, NYC has a large population with relatively high values of travel time who may be more willing to adopt premium and time-saving mobility services such as AAM. Moreover, the city has a long history of operating helicopters for passenger transport, medical flights, and air tours, etc. Therefore, there is both adequate established infrastructure to accommodate such services as well as sufficient market potential with a customer base that would adapt to this type of service.

Airport access is treated as the use case as there is prior research and modeled data available from Rath and Chow (2022), which makes the input parameters more consistent with the literature in the absence of operational data for observation.

The proposed AAM hub system considers a platform that links ride-hail operators with three candidate vertiports, with the research question: how much should an AAM operator of this system subsidize ride-hail/taxi drivers to incentivize them to provide first mile service to passengers to access the vertiports. There might be sufficient demand that no subsidy is necessary, but if costs are too high the lower demand might be worth offsetting with subsidies.

*4.1.1 Data preparation*

The three vertiport candidate locations used in this study are the three current heliports in Manhattan, which are Downtown Manhattan/Wall Street Heliport (JRB), West 30th St Heliport (JRA), and East 34th Street Heliport (6N5), red boxed and shown in Fig. 4. The three airports outlined are Newark Liberty International Airport (EWR), John F. Kennedy International Airport (JFK), and LaGuardia Airport (LGA).

We consider taxi data from 63 taxi zones within Manhattan, obtained from the NYC open data portal (NYC Open Data, 2025). Travel times from JRA (located in NYC's central business district) are typically about 1 hour 25 minutes to JFK, 45 minutes to LGA, and 40 minutes to EWR, whereas an air taxi could complete these trips in typically under 30 minutes, as shown in Table 4. The distribution of business versus leisure trip purposes across the three major airports in the NYC metropolitan are summarized in Table 5 (PANYNJ, 2024).

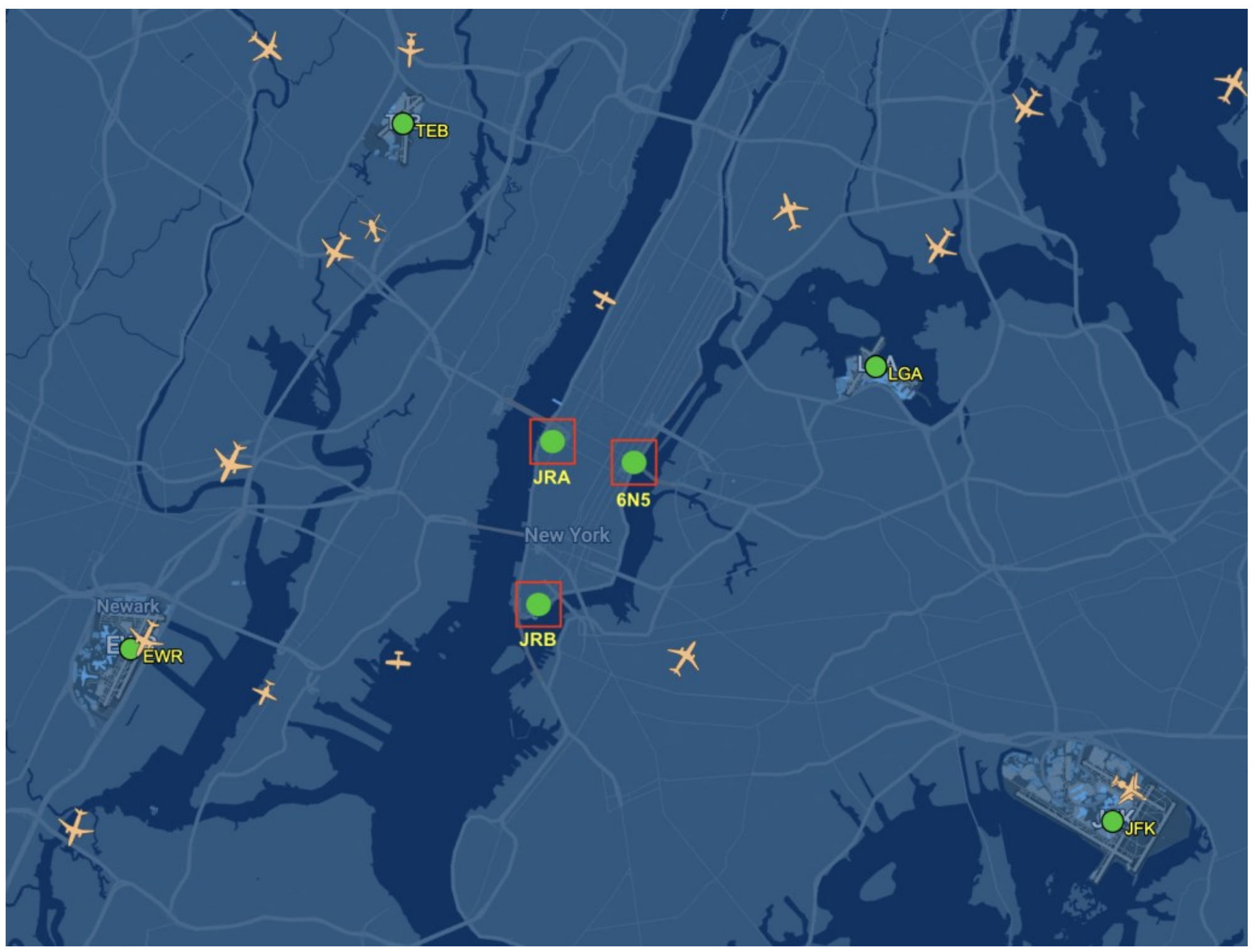

**Fig. 4.** Vertiports and airports used for the case study.

**Table 4**
Air taxi travel time from three vertiports to each airport.

| Origin | Destination | Distance (mi) | Air taxi travel time (min) | Ground taxi travel time (min) |
|---|---|---|---|---|
| JRA | EWR | 14 | 17 | 40 |
| JRA | JFK | 20 | 21 | 85 |

| JRA | LGA | 9 | 15 | 45 |
|---|---|---|---|---|
| 6N5 | EWR | 13 | 17 | 62 |
| 6N5 | JFK | 15 | 18 | 59 |
| 6N5 | LGA | 7 | 14 | 29 |
| JRB | EWR | 7 | 14 | 53 |
| JRB | JFK | 17 | 19 | 79 |
| JRB | LGA | 10 | 16 | 41 |

**Table 5**
Trip purpose distribution for each airport in the NYC area.

| Airport | Business | Leisure |
|---|---|---|
| EWR | 23.5% | 76.5% |
| JFK | 14.0% | 86.0% |
| LGA | 24.2% | 75.8% |

Ground taxi fare is calculated using the high-volume for hire vehicle (FHVHV) monthly trip record data (NYCTLC, 2025a). The data filtering process isolated two different types of trips: 1) outbound trips originating in the Manhattan taxi zones and terminating at one of the three airports, and conversely, 2) inbound trips originating at each airport and terminating in a Manhattan taxi zone. From this retained dataset, the average ground taxi fare per unit distance $f_g$ was computed as the ratio of the aggregate passenger fare to the cumulative travel distance.

The demand data is also obtained from the FHVHV data by filtering the trips to have the origins or destinations that are one of the airports. The processed dataset includes the daily ridership count between different taxi zones and three airports. In using this data as the base demand, the assumption for the AAM hub system is that they are competing primarily with direct taxi trips, since travelers who start from transit or driving are less likely to switch to the more expensive AAM mode. Also, using actual NYC FHVHV data allows us to gauge the spatial demand distribution across different zones to each airport. The road travel distance and travel time (including ground travel times summarized in Table 4) are obtained from Google's distance matrix API.

*4.1.2 Vertiport considerations*

The respective coordinates and aeronautical information of the three candidate vertiports were obtained from FAA (2025). These locations already have operators carrying out air taxi services (i.e. Blade Air Mobility) with helicopters and realistically will be used to accommodate eVTOL air mobility operations if such operations come to NYC. Alvarez et al. (2021) and Rizzi et al. (2024) incorporated the JRB, JRA, and 6N5 heliports as candidate vertiports in their NYC AAM studies to evaluate network-level impacts of integration, using these existing heliports to analyze airspace interactions, delay propagation, system feasibility, and potential acoustic exposure and public response to future eVTOL operations in the city.

*4.1.3 Input parameter assumptions*

All parameters that are used in this case study are listed in Table 6. The VOTs for leisure travelers ($2.25/min) and business travelers ($5.15/min) are adapted from Ma et al. (2017) and Koster et al. (2011), respectively, with inflation adjustments applied to reflect current price levels. For $f_a^{max}$, Blade's current air taxi service airport shuttle price is used (Blade Air Mobility, 2025). Algorithm 1 uses the following hyperparameters: $w_0 = 0.5, h_0 = 1.2, \tau = 0.5, \nu = 0.5, \rho = 1, \eta = 2$, $h_{min} = 5 \times 10^{-6}$, and $W = 2$.

**Table 6**
AAM operations related parameters assumptions.

| Parameter | Definition | Value | Units | Reference |
|---|---|---|---|---|
| $\beta_1$ | VOT of leisure travelers | 2.25 | $/min (Inflation adjusted) | Ma et al. (2017) |
| $\beta_2$ | VOT of business travelers | 5.15 | $/min (Inflation adjusted) | Koster et al. (2011) |
| $c_a$ | Air taxi operator cost | 6 – 14 * | $/mi | Johnston et al. (2020); Garrow et al. (2021) |
| $\kappa$ | Vertiport operating cost | 13 | $/passenger | Butterworth-Hayes (2025) |
| $f_g$ | Ground transport fare | 6.37 | $/mi | NYC Open Data (2025) |
| $f_a^{max}$ | Max air taxi fare | 45.88 * | $/mi | Blade Air Mobility (2025) |
| $c_v$ | Vertiport access operator cost | 14.16 | $/passenger | Parrott & Reich (2019); NYCTLC (2025b) |

Note. Parameter values denoted by an asterisk (*) are estimated under the assumption of a four-passenger air-taxi vehicle.

*4.1.4 Estimation for user cost and operator cost weights*

Recall in the lower-level formulation Eqs. (1) – (6), there are $\alpha_1$ and $\alpha_2$ in place to scale the user cost and operator cost term respectively. To estimate these parameters, we adopt the estimation method in Fosgerau et al. (2022). Since operational flow observations are not available, we refer to a prior study of air taxi airport access in NYC (Rath & Chow, 2022), where a mode choice model was specified. We use the model to synthesize flows in our case study network from which the parameters are then estimated. In effect, this estimation effort maps the utility functions in the model from (Rath & Chow, 2022) over to the PURC-based traveler-operator coalitional choice model specification.

Given the synthesized flows, link lengths, and node-link incidence, we construct the node-link incidence matrix $A$ and the link selector identity matrix $B$. We then compute the generalized inverse $C$ using the Moore–Penrose pseudoinverse of $BA^T$. With this setup, we define the regression variable $y_s$ as shown in Eq. (14). The regression variable $w_s$ is defined in Eq. (15) which captures the operator-related costs. Here, $z_s$ is a matrix encoding the attributes of the utility function that are operator-specific, with a value of 1 for distance-based attributes in this case. Together, $y_s$ and $w_s$ form the explanatory variables representing user and operator cost sensitivities across OD pairs.

$$y_s = (I - B_s A^T C_s) B_s (l \circ F'(\hat{x}_s)) \tag{14}$$

$$w_s = (I - B_s A^T C_s) B_s (l \circ z_s) \tag{15}$$

Eq. (16) specifies the regression equation allowing us to estimate the weights $\alpha_1$ and $\alpha_2$. The estimation is performed using ordinary least squares (OLS), with results summarized in Table 7.

$$y_{si} = \alpha_1 W_{user} + \alpha_2 W_{operator} + \epsilon \tag{16}$$

**Table 7**
OLS estimation results

| Predictor | Coefficient | Std. Error | p |
|---|---|---|---|
| $\alpha_1$ | 0.6068 | 0.0006 | < .001 |
| $\alpha_2$ | 0.1019 | 0.00004 | < .001 |

The OLS results (Table 7) estimates indicate a good fit with $R^2$ = 0.921. The coefficients $\alpha_1 > \alpha_2$ imply that user cost sensitivity has a greater effect on the model, confirming that passengers strongly respond to perceived travel costs.

## 4.2 Results

### *4.2.1 Model results*

The numerical experiments for the case study were performed on a laptop using Python 3.11 with an Apple M3 CPU and 16 GB of RAM. In the model, air operating cost $c_a$ constitutes a main component of the operator's expenditure and therefore serves as a key parameter for reference. Therefore, conducting sensitivity analysis of $c_a$ is necessary for understanding implications on profitability and related decision variables. The sensitivity analysis in Table 8 shows how changes in $c_a$ can influence other variables. The columns corresponding to "Without subsidy" refer to solving the problem while forcing subsidy to be zero, representing a benchmark operation without the AAM hub platform design allowing integration of subsidized ride-hail access.

**Table 8**
Sensitivity analysis for $c_a = [10, 15]$ $\$/mi$.

| | Without subsidy | | | With subsidy | | | | |
|---|---|---|---|---|---|---|---|---|
| $c_a$ (\$/mi) | $\tilde{f}_a$ (\$/mi) | $\tilde{v}_l$(pax) [% share] | $\tilde{\Omega}_0$ (\$) | $f_a$ (\$/mi) | $v_l$ (pax) [% share] | $\Omega_0$ (\$) | $r$ (\$/mi) | $\mathcal{R}$ (\$) |
| 10 | 13.58 | 3551[10.84] | $7.42 \times 10^4$ | 13.58 | 3551[10.84] | $7.42 \times 10^4$ | <0.01 | N/A |
| 10.5 | 13.84 | 2838[8.66] | $5.90 \times 10^4$ | 13.84 | 2838[8.66] | $5.90 \times 10^4$ | <0.01 | N/A |
| 11 | 14.06 | 2253[6.87] | $4.49 \times 10^4$ | 14.06 | 2251[6.87] | $4.49 \times 10^4$ | <0.01 | N/A |
| 11.5 | 14.93 | 1475[4.50] | $3.45 \times 10^4$ | 14.92 | 1472[4.49] | $3.45 \times 10^4$ | <0.01 | N/A |
| 12 | 15.86 | 862[2.63] | $2.44 \times 10^4$ | 15.65 | 985[3.01] | $2.83 \times 10^4$ | 1.66 | $6.52 \times 10^3$ |
| 12.5 | 16.24 | 707[2.16] | $1.91 \times 10^4$ | 16.01 | 837[2.56] | $2.22 \times 10^4$ | 1.75 | $6.19 \times 10^3$ |
| 13 | 16.71 | 572[1.75] | $1.48 \times 10^4$ | 16.43 | 684[2.09] | $1.73 \times 10^4$ | 1.83 | $5.28 \times 10^3$ |
| 13.5 | 16.96 | 485[1.48] | $1.11 \times 10^4$ | 16.65 | 566[1.73] | $1.31 \times 10^4$ | 1.86 | $4.07 \times 10^3$ |
| 14 | 17.42 | 379[1.16] | $8.27 \times 10^3$ | 17.10 | 446[1.36] | $9.73 \times 10^3$ | 1.87 | $3.28 \times 10^3$ |
| 14.5 | 17.60 | 306[0.93] | $6.17 \times 10^3$ | 17.26 | 359[1.10] | $7.27 \times 10^3$ | 1.93 | $2.76 \times 10^3$ |
| 15 | 17.88 | 227[0.69] | $4.62 \times 10^3$ | 17.52 | 286[0.87] | $5.46 \times 10^3$ | 1.99 | $2.47 \times 10^3$ |

When $c_a$ increases from \$10/mi to \$15/mi, the objective value decreases from \$74,205.41 to \$5,459.28, showing the air taxi operational profit declines as the operating costs increase. The air taxi fare $f_a$ also increases with $c_a$, as enforced by Eq. (8), which ensures that the air-taxi operator does not operate at a deficit.

**Remark 1**. *The model identified the threshold in air operating cost over which the AAM provider has incentive to subsidize ride-hail first mile providers.*

To examine changes in subsidy, it can be observed that when the air-taxi operating cost remains $c_a \in [10,12]\$/mi$, the optimal subsidy rate $r^*$ is effectively negligible. However, once the air-taxi operating cost $c_a$ exceeds \$12/mi, the optimal subsidy increases to \$1.66/mi, indicates a transition from a zero subsidy corner solution to a positive subsidy interior optimum. This change reflects a transition in the upper-level optimization from a corner solution, in which a zero subsidy is optimal because the benefit of subsidizing vertiport access is insufficient to offset its cost. At this threshold, the fare increases required to maintain nonnegative profitability induce a sufficiently large reduction in demand that subsidizing vertiport access yields net gains by partially offsetting generalized access costs. The gain is approximately in the range of 16% increase in the upper-level objective value.

**Remark 2**. *The model quantified the upper air operating cost bound $\underline{c}_a^{leisure}$ ($\underline{c}_a^{business}$) above which it is not profitable to serve leisure (business) customers; $\underline{c}_a^{leisure} < \underline{c}_a^{business}$.*

Fig. 5 depicts the relationship between the upper-level objective value and the air-taxi operating cost $c_a$for two passenger segments differentiated by their VOT. As $c_a$increases, the overall upper-level objective value (solid green line) declines monotonically, reflecting the need to raise the air-taxi fare $f_a$ to offset higher operating costs. Although both passenger segments exhibit this downward trend, the magnitude of the decline differs across groups. In particular, the business passenger segment (blue dash-dotted line, $\beta_1 = \$5.15/\text{min}$) experiences a slower reduction in objective contribution than the leisure segment (red dashed line, $\beta_2 = \$2.25/\text{min}$), despite representing a smaller share of the overall population.

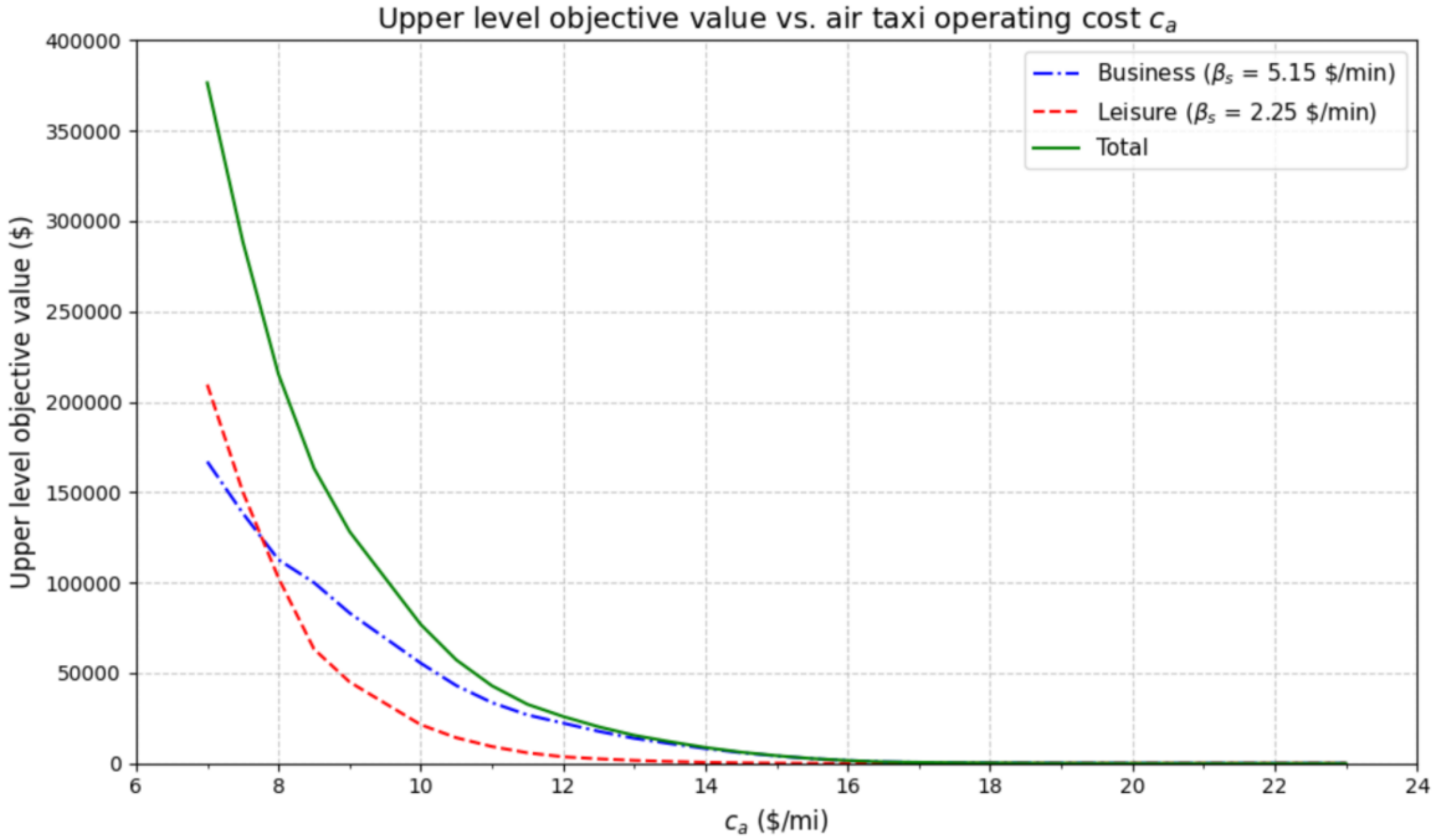

**Fig. 5.** Upper-level objective value as $c_a$ increases, shown separately for business, leisure, and total passenger demand.

**Remark 3**. *The model identified the upper air operating cost bounds* $\underline{c}_a^{s,vertiport}$ *above which specific vertiports are not profitable to include in the AAM hub platform; the ranking of these vertiports depends on the network and demand structure.*

The air taxi passenger ridership at each vertiport is also obtained from the lower-level model results. Fig. 6 shows that ridership declines consistently with increasing $f_a$ across all vertiports, indicating a consistent negative price-demand relationship. At any given fare level, the demand curves for business travelers (dash-dotted lines) lie uniformly above those for leisure travelers (dashed lines), reflecting the higher willingness to pay of passengers with greater VOT. It is not immediately obvious which vertiport would have the lowest upper cost bound $\underline{c}_a^{s,vertiport}$. As shown in Fig. 6, initially JRB has lowest number of passengers at $c_a = \$7$, but JRA is the first vertiport to reach upper bound $\underline{c}_a^{s,JRA}$ for both segments. Because the vertiports have overlapping catchment areas, when costs go up they may cannibalize one another. This result argues for the need for model analysis.

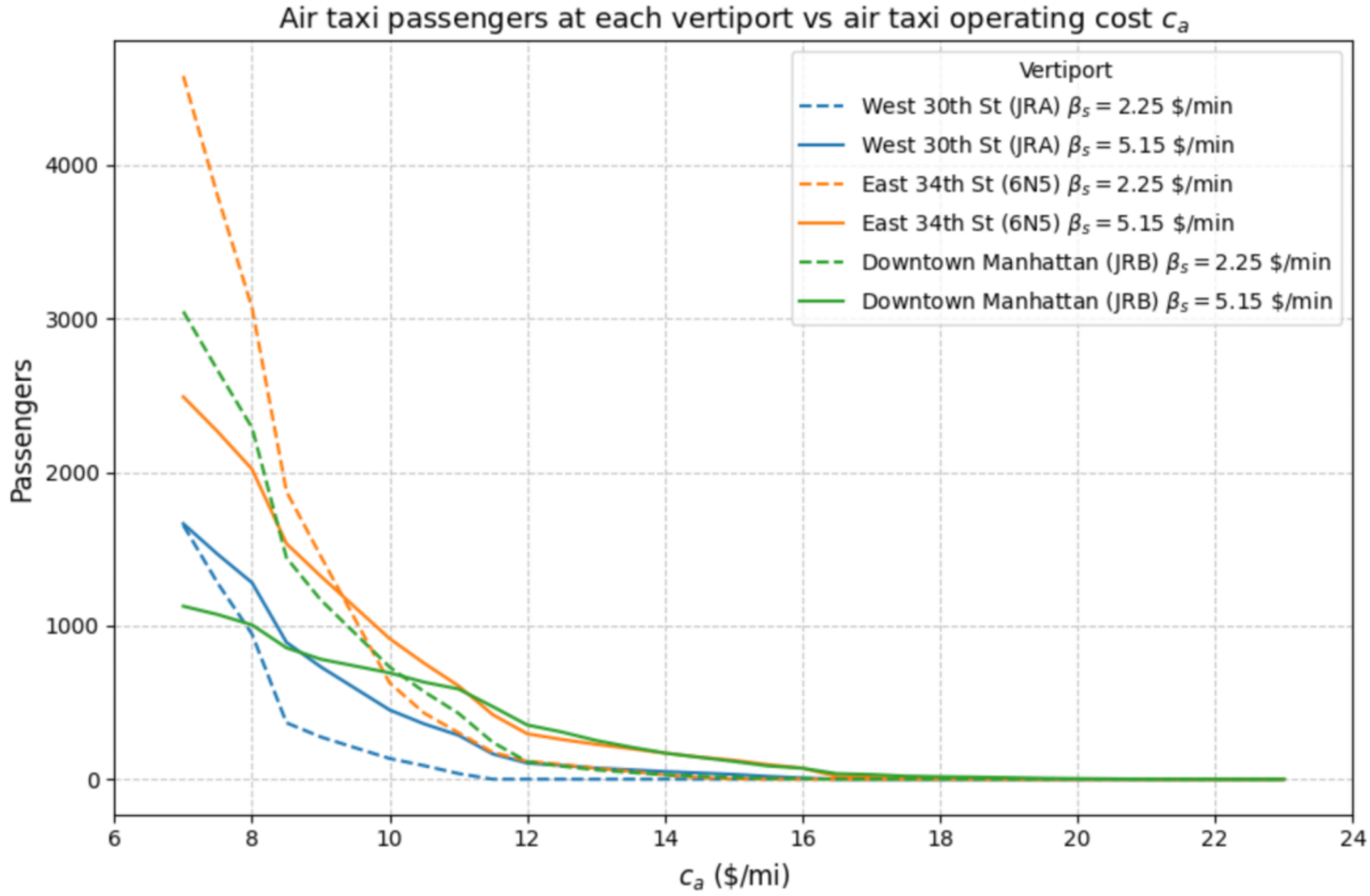

**Fig. 6.** Effect of air-taxi operating cost $c_a$ on air taxi ridership composition by vertiport and passenger segment.

Fig. 7 presents the distribution of air-taxi ridership at each vertiport and highlights the relative shares of two passenger segments via their corresponding pie charts. In the $c_a = \$8/mi$ case, the air-taxi operating cost is comparatively low, resulting in a larger share of leisure passengers at all three vertiports (42.5% at JRA, 60.3% at JRB, and 69.5% at 6N5). However, as operating cost increases, the leisure segment declines as air taxis become less competitive relative to the direct ride hailing alternative. These results further indicate that once the operating cost reaches $\$12/mi$, the subsidy mechanism begins to impact air taxi ridership composition at different vertiports. Notably, at $c_a = \$12/mi$, 6N5 accommodates fewer passengers (413) compared with JRB (467), yet the total subsidy amount 6N5 is larger due to longer average ground access distances. This occurs because 6N5 attracts passengers from taxi zones in upper Manhattan, whereas JRB primarily serves passengers originating from closer taxi zones.

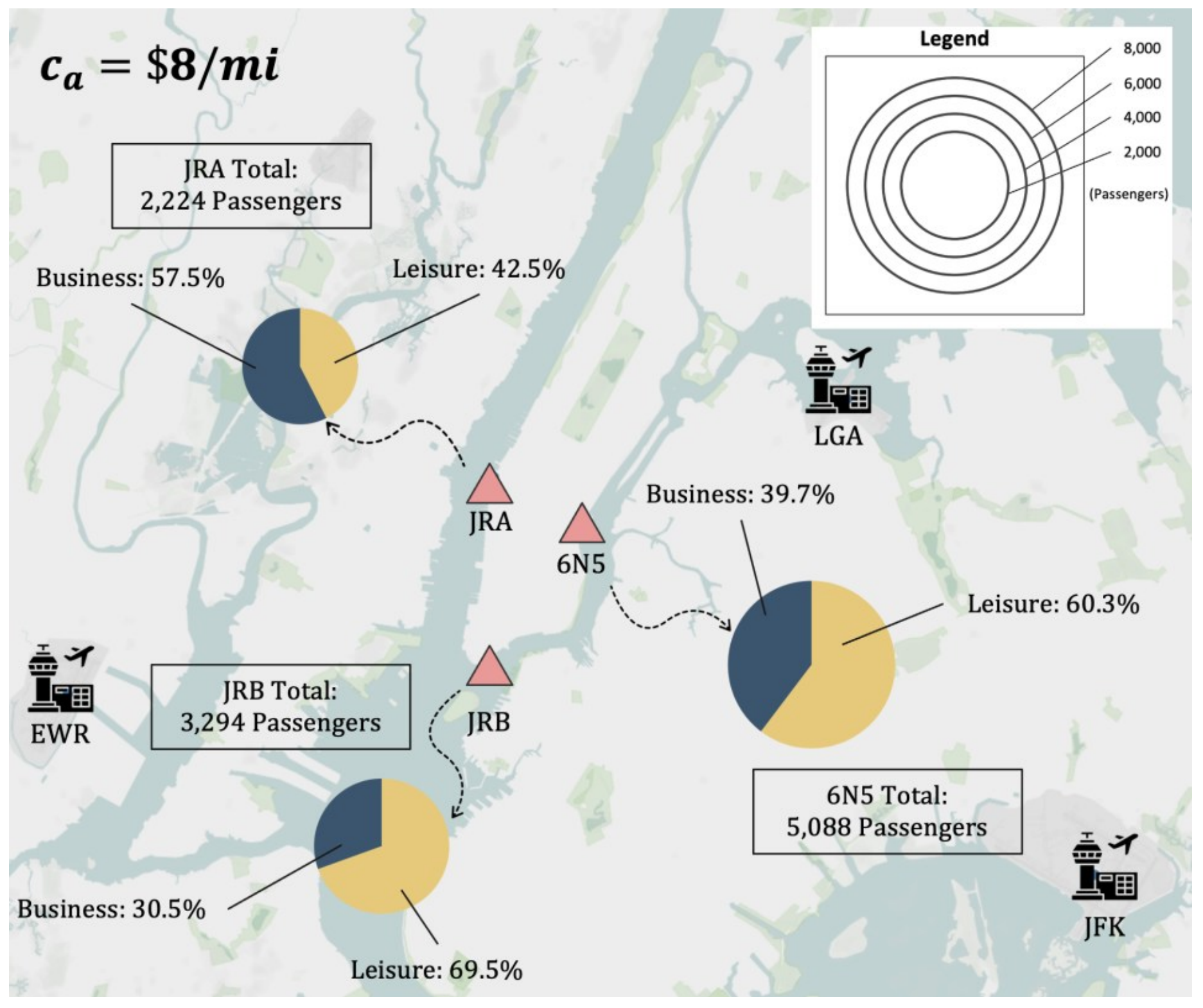

$c_a = \$8/mi$
Legend
8,000
6,000
4,000
2,000
(Passengers)
JRA Total:
2,224 Passengers
Business: 57.5%
Leisure: 42.5%
LGA
JRA
6N5
Business: 39.7%
Leisure: 60.3%
EWR
JRB Total:
3,294 Passengers
JRB
6N5 Total:
5,088 Passengers
JFK
Business: 30.5%
Leisure: 69.5%

(a)

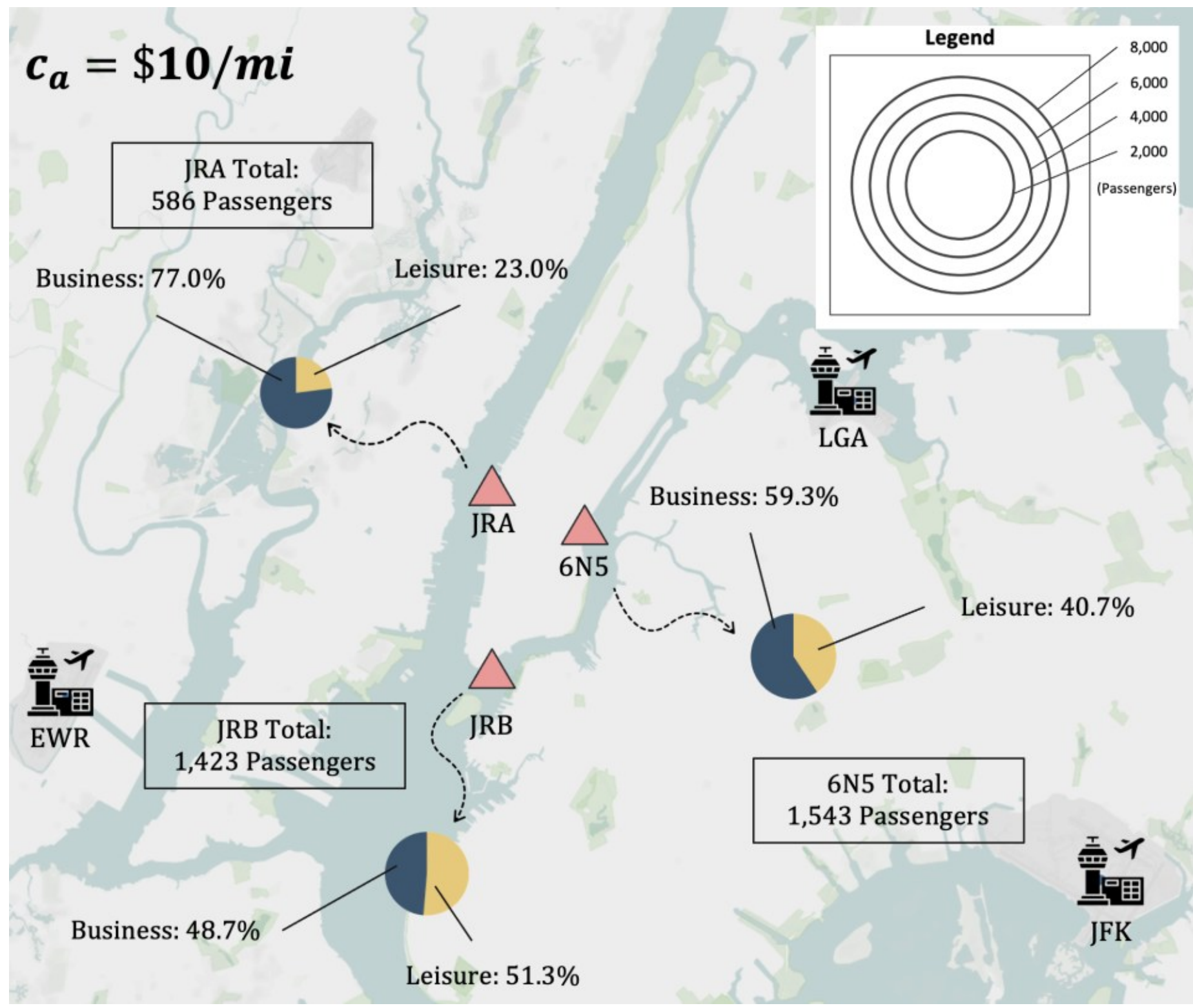

$c_a = \$10/mi$
Legend
8,000
6,000
4,000
2,000
(Passengers)
JRA Total:
586 Passengers
Business: 77.0%
Leisure: 23.0%
LGA
JRA
6N5
Business: 59.3%
Leisure: 40.7%
EWR
JRB Total:
1,423 Passengers
JRB
6N5 Total:
1,543 Passengers
JFK
Business: 48.7%
Leisure: 51.3%

(b)

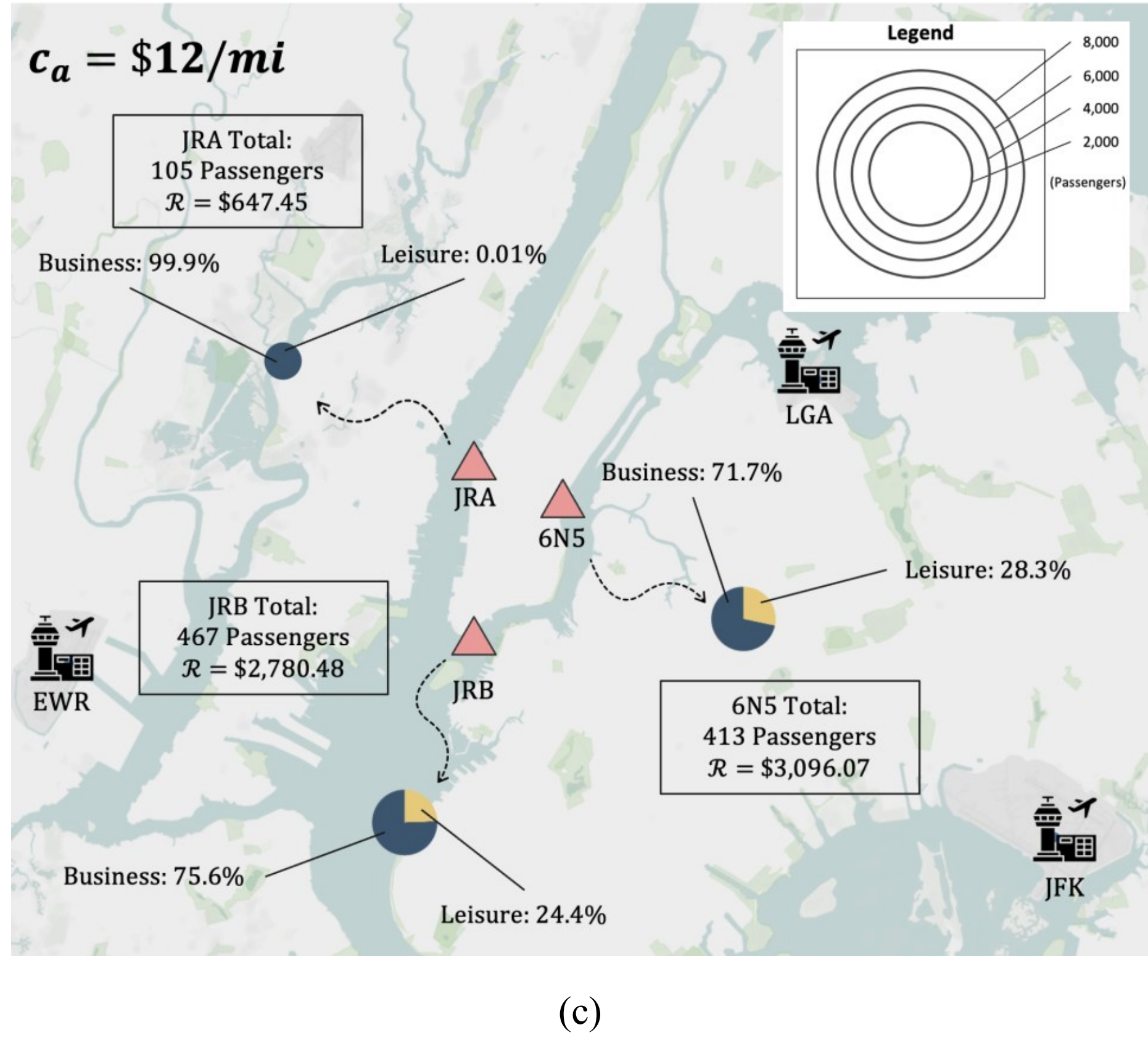

(c)

**Fig. 7.** Distribution of air taxi demand by trip purpose at each vertiport: (a) $c_a = \$8\ /mi$; (b) $c_a = \$10\ /mi$; (c) $c_a = \$12/mi$.

**Remark 4**. *The ranking of relative importance of each vertiport to the AAM platform upper level objective depends on the air taxi operating cost.*

Fig. 8 examines the contribution of each vertiport to the upper-level objective by evaluating how the objective value changes when individual vertiports are removed across different $c_a$. At lower air taxi operating costs ($at\ c_a = \$8\ /mi$), the removal of 6N5 generates the largest reduction in the upper-level objective (47.98%), followed by JRB (31.06%) and JRA (20.97%). As $c_a$ increases, the contribution of JRA decreases monotonically, while JRB's contribution increases and eventually surpasses that of 6N5 (at $c_a = \$10.5\ /mi$), becoming the vertiport with the highest contribution to the upper-level objective.

These differences reflect spatial demand heterogeneity within Manhattan and each vertiport's relative importance to the upper-level objective. By quantifying the loss in profit associated with vertiport removal, Fig. 8 provides a basis for prioritizing vertiport investments. The results suggest that investments in JRB and 6N5 yield the greatest improvements in the operator's profit, whereas investments in JRA are advantageous primarily when $c_a$ is relatively low.

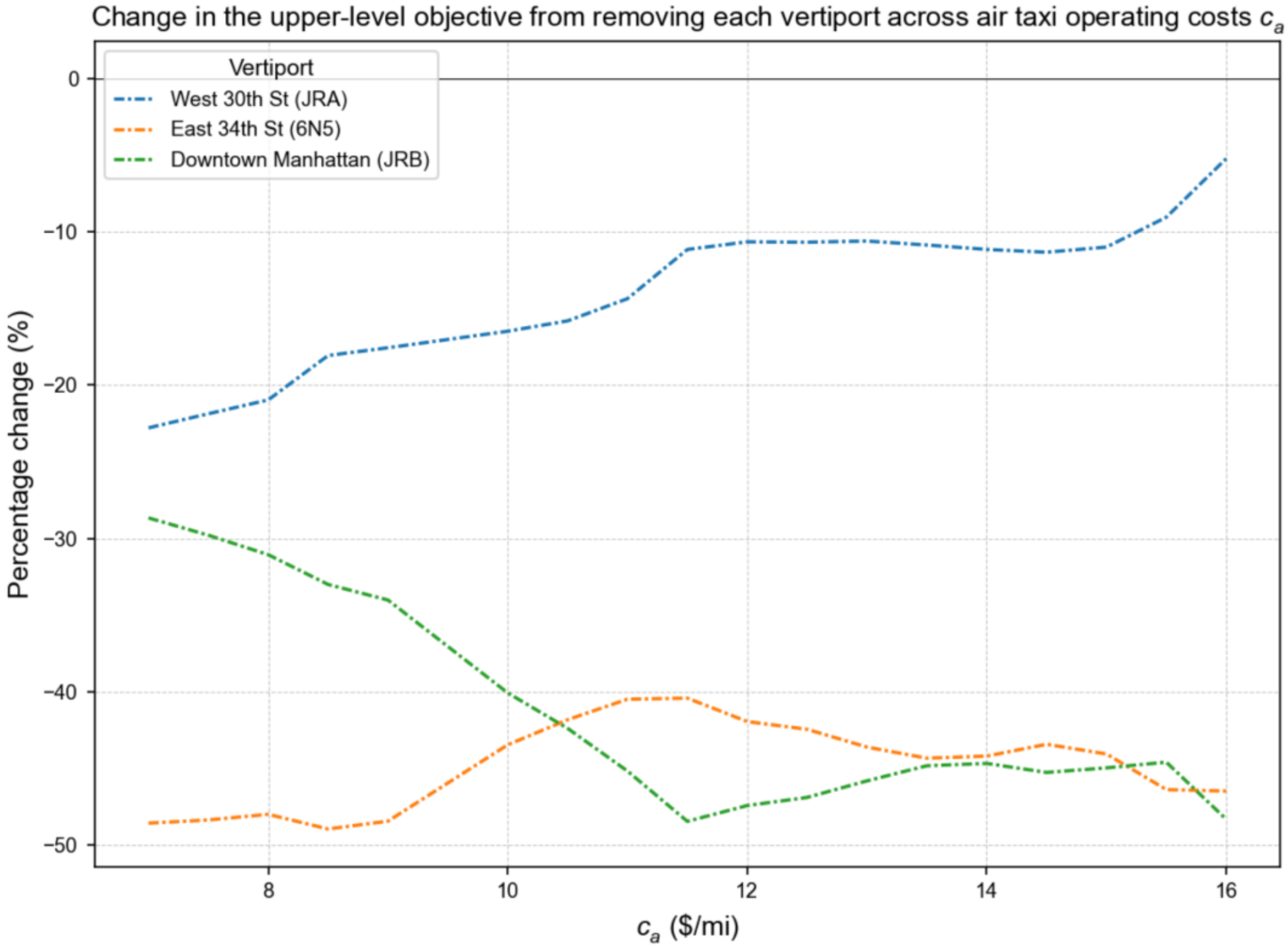

**Fig. 8.** Percent change in upper-level objective with each vertiport's removal under different air taxi operating cost $c_a$.

## 5. Conclusion

The study proposes a framework for AAM multimodal operations that can determine the optimal distance-based air taxi fare and subsidy rate to incentivize travelers to use ride-hailing service to access the vertiports. The proposed model can capture the interactions between the operators and travelers, offering a realistic decision-support tool for AAM operations, where such multi-operator trips can be booked and coordinated through digital platforms. The proposed method was applied to the NYC airport-access network as a case study, utilizing real-world demand and geospatial data to capture realistic travel behavior.

In summary, the results indicate that once air-taxi operating costs exceed approximately $12/mi, access subsidies become a necessary instrument for sustaining AAM ridership by offsetting higher generalized access costs. Among the three candidate vertiports, 6N5 has the highest contribution to upper level objective value especially at lower operating costs, followed by JRB and then JRA. However, as $c_a$ increases, JRB's impact increases and eventually surpasses 6N5, while JRA plays a comparatively smaller role. In addition, despite representing a smaller share of the passenger population, business travelers emerge as the most viable passenger segment due to their higher VOT and lower sensitivity to fare increases. Although the case study focuses on airport-access AAM services, the proposed modeling framework is transferable to other multimodal AAM applications. Future research may extend this work by integrating fleet scheduling and charging operations to examine how fleet availability, scheduling and energy constraints jointly influence optimal pricing and subsidy decisions under varying demand and cost scenarios.

## Acknowledgments

Funding support from NSF CMMI-2423908 is gratefully acknowledged. Zhenglei Ji was also partially supported by the Dwight David Eisenhower Transportation Fellowship Program (DDETFP) Graduate Fellowship awarded by the United States Department of Transportation.